\def\VEV#1{\left\langle\,#1\,\right\rangle}
\def\rf#1{(\ref{#1})}

\documentclass[aps,prl,twocolumn]{revtex4}

\usepackage{amsmath}
\usepackage{graphicx}

\begin{document}

\title{Tracer diffusion in a dislocated lamellar system}

\author{Victor Gurarie}
\affiliation{Department of Physics, Theoretical Physics, 1 Keble Road,
Oxford University, Oxford OX1 1JP, UK}

\author{Alexander E.~Lobkovsky}
\affiliation{Physics Department, Northeastern University, 110 Forsyth
  Street, Boston, MA 02115}




\date{\today}


\begin{abstract}
  Many lamellar systems exhibit strongly anisotropic diffusion.  When
  the diffusion across the lamellae is slow, an alternative mechanism
  for transverse transport becomes important.  A tracer particle can
  propagate in the direction normal to the lamellae, never leaving a
  particular layer, by going around a screw dislocation.  Given the
  density of positive and negative screw dislocations, we calculate
  the statistical properties of the transverse transport.  When either
  positive or negative dislocations are in excess, the tracer moves
  ballistically normally to the layers with the mean square of the
  displacement growing like the square of time $T^2$.  When the
  average dislocation charge is zero, the mean square of the normal
  displacement grows like $T \log T$ for large times.  To obtain this
  result, the trajectory of the tracer must be smoothed over distances
  of order of the dislocation core size.
\end{abstract}

\maketitle

Diffusion in layered systems is often strongly anisotropic.  The
mechanisms and the manifestations of the anisotropy vary.  In many
such systems diffusion across is significantly slower than along the
lamellae.  For example, enhanced creep resistance of lamellar alloys,
such as industrial TiAl, is probably due to the high barrier for the
dislocations crossing from one layer to the next
\cite{johnson00:_creep_tial_si,wang98:_tial_ledges}.  In lamellar
phases of diblock copolymers \cite{fredrickson96:_copolymers}, tracer
diffusion along the lamellae can be up to forty times faster than
across \cite{hamersky98:_anisotropy}.  The fact that water diffusion
in lamellar phases of phospholipid bilayers
\cite{wassall96:_pulsed_nmr} is strongly anisotropic may be relevant
to attempts to use multilamellar vesicles for drug delivery
\cite{demenezes96:_study}.  Another example of anisotropic diffusion
is the kinetics of electroactive probes in lyotropic liquid crystals
\cite{postlethwaite94:_elect_cesium}.

When lateral diffusion is much faster than transverse diffusion, the
tracer can still be transported quickly in the direction normal to the
lamellae if screw dislocations are present in the system.  A screw
dislocation is constructed by cutting a perfect layered structure with
a half-plane normal to the layers, shifting the two sides of the cut
with respect to each other in the direction normal to the layers by a
distance equal to the layer spacing, and finally gluing the cut.
Screw dislocations are indeed common in a variety of layered systems
\cite{allain86:_possib,dhez00:_lamel,j.petermann79:_obser}.  A summary
of various dislocation properties in lamellar systems is presented in
Ref.~\cite{holyst95:_disloc}.  

When a tracer particle confined to a particular layer encircles a
screw dislocation, it finds itself in one layer higher (or lower).  A
tracer particle can then reach any point in the system while remaining
confined to a layer.  The trajectory of the tracer projected onto a
plane parallel to the layers is a two-dimensional random walk.  Upon
completing a closed 2D trajectory, our random walker moves up or down
the number of layers equal to the dislocation charge enclosed by the
trajectory.  We obtain an expression for this quantity by noting that
when a single screw dislocation is present in the system, the layer
number of the walker is the winding angle around the dislocation
divided by $2\pi$.  Consider now the sum of the winding angles around
all the dislocations (the signs of the individual winding numbers are
determined by the charge of the dislocations).  This quantity changes
continuously.  The change in this quantity along an open trajectory
depends on the shape of the sample due to the contributions of the
winding numbers around distant dislocations.  However, when the walker
returns to the origin, the change in the total winding number is the
dislocation charge enclosed by the trajectory.  Thus we identify the
total winding number divided by $2\pi$ with the layer number $n(t)$
which for closed trajectories coincides with the normal displacement
of the walker.

In this letter we study the diffusion of a tracer particle confined to
the lamellae.  The tracer starts at the origin of layer $n = 0$ at
time $t = 0$ and explores the $x$-$y$ plane with diffusivity $D$.  Let
there also be a random distribution of positive and negative screw
dislocations with densities $f_+$ and $f_-$ respectively.  Our goal is
to determine the nature of the transport normal to the layers by
predicting the result of the following experiment.  If some amount of
the tracer material is placed at the origin at time $t = 0$, what is
the density of the resulting cloud of tracer particles as a function
of time?

\begin{figure}[htbp]
  \includegraphics[width=6cm,keepaspectratio]{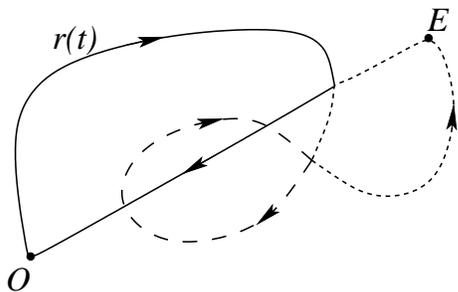}
  \caption{We complete the path $r(t)$ with a straight segment
    connecting its beginning $O$ and its end $E$.  The winding angle
    fluctuation along this closed path can be calculated by
    decomposing the path into a union of non self-intersecting loops
    (denoted by the solid, dashed and dotted lines).}
  \label{fig:combined}
\end{figure}

To accomplish this task we look at paths which start at the origin $O$
(see Fig.~\ref{fig:combined}) at time $t = 0$ and arrive to point $E$
located a distance $R$ from the origin at time $T$.  We seek to define
the layer number $n(R, T)$.  Since the layer number change is only
well defined for closed trajectories we fix $n(R, T)$ by completing
the path $r(t)$ with a straight segment $OE$ connecting this point to
the origin.  We can then define $n(R, T)$ to be the total dislocation
charge enclosed by this trajectory.

We now seek to average powers of $n$ over positions of dislocations
and random walks which end at $r = R$ at time $t = T$.  We denote the
average over positions of dislocations with an overbar and average
over random walks by angular brackets $\VEV{\cdot}$.  Because changing
the shape of the completing segment adds a constant to $n(R, T)$, its
average $\VEV{\overline n(R, T)}$ has no physical meaning.  However,
its standard deviation $\sigma(R, T) = \VEV{\overline{n^2}} -
\VEV{\overline n}^2$ is independent of the shape of the completing
segment.  It gives the size of the spreading tracer cloud at time $T$
and distance $R$ from the origin.

We identified two qualitatively different cases.

When $f_+ \not = f_-$, we are able to obtain $\sigma(R, T)$ thereby
predicting the tracer density profile within the spreading cloud.
This is possible because the average total dislocation charge within a
closed trajectory which is proportional to its signed area.  In this
case we find that the vertical size of the tracer cloud grows linearly
in time, i.e.\ there is superdiffusion across the layers.  Moreover,
the spreading cloud acquires a biconcave shape, since $\sigma(R, T)
\propto D^2 T^2 + 2 R^2 D T$ (here brackets denote averaging over
random walks).  We must note here that the excess of dislocations of a
certain chirality leads in smectics to the break up of the homogeneous
lamellar phase into domains separated by twist grain boundaries
\cite{bluestein01:_disloc_tgb}.  We nevertheless pursue this case
since it may applicable to the Aharonov-Bohm electron phase
fluctuations in a type II superconductor and other systems where
geometric winding numbers play a role.

When the densities of the positive and negative dislocations are
equal, a more subtle averaging must be performed since the average
dislocation charge in a closed loop is now zero.  We must compute the
variance of the dislocation charge within a closed loop.  As we
discuss below, this variance is proportional to the unsigned area of
the loop which has a simple geometrical interpretation.

It turns out that when the dislocations are thought of as point
objects, it is not possible to average the unsigned area over random
walks.  The variance of the layer number obtained in this fashion is
logarithmically divergent.  Divergences are common in the statistics
of winding numbers of random walks
\cite{spitzer58:_some,pitman86:_asymp,drossel96:_winding,kholodenko98:_elemen_brown}.
For example, the dispersion of the winding number of a random walker
around a point is divergent if the walk is continuous.  This
divergence arises due to the contributions to the winding number from
trajectories which wind tightly around the point.  The nature of our
divergence is similar.  By traversing a short distance around a
dislocation a tracer particle travels far in the direction normal to
the layers, and that leads to the anomalously fast diffusion in the
direction perpendicular to the layers.

We regularize this divergence by noting that the core size $a$
determines the distance of the closest approach of the tracer to the
dislocations.  Therefore small loops in the trajectory are irrelevant
for our purposes.  We should therefore look at an effective discrete
random walk whose steps are of length $a$ taken every $a^2/D$ seconds.
We then are able to calculate the variance of the layer number which
grows as $\sigma(R, T) \propto T \log T$.  Since $\sigma(R, T)$ is
independent of $R$,, the shape of a spreading cloud in this case is an
ellipsoid which elongates in the direction normal to the layers.  Note
also that this divergence would have appeared in the case of different
densities of positive and negative dislocations.  It leads to a
correction of order $T \log T$.

We now describe our methods and results in more detail.  Let
$r_\alpha(t)$ be the Brownian trajectory of the tracer (here $\alpha$
is the two dimensional vector index).  We take its velocity $\dot
r_\alpha(t)$ to be random and white noise correlated in time,
neglecting possible correlations on time scales of the order of the
scattering time of the walker, which is much smaller than all other
time scales in our problems.
\begin{equation}
  \label{eq:vcor}
  \VEV{\dot r_\alpha (t) \dot r_\beta(t')} = \delta_{\alpha \beta} 
  \left[ {D \over 2}  \left( \delta(t - t') -{1 \over T} \right) +
    {R_\alpha^2 \over T^2} \right].
\end{equation}
Eq.~(\ref{eq:vcor}) involves constant terms in addition to the
standard $\delta(t - t')$ one.  This is because averaging in
Eq.~(\ref{eq:vcor}) is done with the boundary conditions $r(0) = 0$,
$r_\alpha(T) = R_\alpha$, needed to compute the layer number $n$ as a
function of position $R$ and time $T$.

Dislocations of charge $q_i$ are located at $x_\alpha^i$.  $q_i$ takes
on values $\pm 1$.  The layer number $n(R, T)$ can be expressed in the
following way
\begin{equation}
  \label{eq:ang}
  n(R,T) = \sum_i {q_i \over 2 \pi} \int dt \ {\epsilon_{\alpha
      \beta} \dot r_\beta \left(r_\alpha - x_\alpha^i \right) \over
      |r - x^i|^2},
\end{equation}
where $\epsilon_{\alpha\beta}$ is the antisymmetric tensor of rank 2.
Indeed, the expression to be integrated over time in
Eq.~(\ref{eq:ang}) is just the sum over all the dislocations of the
infinitesimal change of the angle between the $x$-axis and the vector
connecting the tracer particle and the dislocation.  Thus
(\ref{eq:ang}) is the cumulative winding number of the tracer around
the dislocations.  According to the preceding discussion (see
Fig.~\ref{fig:combined}), the function $r(t)$ in \rf{eq:ang} consists
of two segments.  The first is a Brownian walk from $t = 0$ to time
$T$.  The second is a straight line from $r(T) = R$ back to the
origin.

Since we are interested in the statistical properties of $n(R, T)$,
expression (\ref{eq:ang}) must be first averaged over dislocation
strengths $q_i$ and positions $x_i$ and then over Brownian
trajectories $r(t)$.  To perform the averaging, we assume that
positive and negative dislocations are distributed uniformly.  If the
total density is $f = f_+ + f_-$, the dislocation strength is $q_i =
1$ with probability $f_+/f$, and $q_i = -1$ with probability $f_-/f$.

Averaging \rf{eq:ang} over the strengths and positions of the
dislocations we arrive at
\begin{equation}
  \label{eq:ang_q_i_position}
  \overline n(R,T) = (f_+ - f_-) \int dt \ \frac{\epsilon_{\alpha
      \beta}}{2}\, r_\alpha \dot r_\beta.
\end{equation}
The integral in (\ref{eq:ang_q_i_position}) can be interpreted
geometrically as the overall area covered by a vector connecting the
tracer particle to the origin as the particle moves along its
trajectory.  The area is computed with the sign, so that when the
vector rotates clockwise, the area it covers is added, while when it
moves counterclockwise, it is subtracted from the answer. We refer to
the integral in \rf{eq:ang_q_i_position} as the signed area.

Eq.~(\ref{eq:ang_q_i_position}) has a simple intuitive interpretation.
If the densities of the positive and negative dislocations were the
same, we would expect $\overline n(R, T)$ to vanish, because on
average the tracer would encircle an equal number of positive and
negative dislocations.  The signed area times the difference in the
dislocation densities is on average precisely the overall number of
dislocations encircled clockwise minus the overall number of
dislocations encircled counterclockwise, which should give $\overline
n(R, T)$.

\begin{figure}[htbp]
  \centering
  \includegraphics[width=6cm,keepaspectratio]{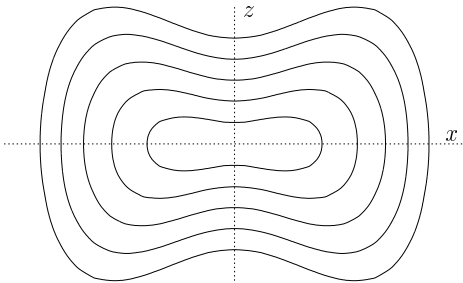}
  \caption{Five equidistant in time snapshots of the isodensity line
    of the vertical slice through the expanding tracer cloud when $f_-
    \not = f_+$.  The cloud's shape is a figure of revolution of this
    slice around the $z$-axis.  The lines are drawn at the level
    where the density is equal $0.1$ of the maximum density in the
    center of the cloud.  Note that that cloud develops into a
    biconcave shape elongated in the vertical direction (normal to the
    layers).}
  \label{fig:cloud}
\end{figure}

At this point, we must clearly distinguish between the $f_- = f_+$ and
$f_- \not = f_+$ cases.  If the difference in dislocation densities is
zero, then $\overline n(R, T)$ computed in this way is zero and we
must take into account the fluctuations in the total dislocation
charge enclosed by a trajectory.  Let us first concentrate on the case
$f_- \not = f_+$.

With these preparations, it is now straightforward to average powers
of $\overline n(R, T)$ over the Brownian walks, with the help of
\rf{eq:vcor}.  The average $\VEV{\overline n}$ of
(\ref{eq:ang_q_i_position}) can be shown to be zero, owing to the
straight shape of the completing segment $OE$ of
Fig.~\ref{fig:combined}.  The standard deviation $\sigma(R, T)$ can
then be computed as $\overline{n^2}(R, T)$.  However we can neglect
the difference $\VEV{\overline{n^2}} - \VEV{\overline n^2}$ which can
be shown to grow slower in time than $\sigma(R, T)$.  We obtain
\begin{equation}
  \label{eq:nn}
  \sigma(R, T) \approx \VEV{\overline n^2} = \frac{(f_+ - f_-)^2}{48}
  \left( D^2 T^2 + 2 R^2 DT
  \right).
\end{equation}
This is the first of the two main results of this letter.  Noting that
$\VEV{R^2} = DT$ for a Brownian walker, we conclude that the tracer
particle indeed moves superdiffusively in the normal direction.
Furthermore, \rf{eq:nn} gives us a way to calculate the approximate
shape of a tracer cloud shown in Fig.~\ref{fig:cloud}.

Moreover, it is also possible to compute the entire probability
distribution of $\overline n(R, T)$ which allows us to determine the
density of the cloud.  This calculation involves averaging the
exponential of \rf{eq:ang_q_i_position} over the Brownian walks using
Gaussian functional integral techniques.  The answer, given in terms
of infinite products, is not illuminating.  We only note here that the
probability distribution $P(\overline n, T)$ of a simpler quantity
$\overline n(T)$, which is the average of $\overline n(R, T)$ over all
positions $R$, can be calculated in closed form,
\begin{equation}
  P(\overline n, T) = {2 \over \left|f_+ - f_- \right| D T}
  \left[
    \cosh \left( {2 \pi \overline n \over \left(f_+ - f_- \right) D T
  }\right)
  \right]^{-1}.
\end{equation}

The situation becomes more interesting when $f_- = f_+$.  In this
case, to compute $\overline{n^2}$ we need to square \rf{eq:ang} first
and then average over positions and strengths of dislocations.  We
obtain
\begin{equation}
  \label{eq:unsign}
  \overline{n^2}(R, T) = - {f \over 4 \pi} \int dt \int dt'~\dot
  r_\alpha(t) \, \dot r_\beta(t') \, \,G_{\alpha \beta}(r(t) - r(t')),
\end{equation}
where $G_{\alpha \beta}(r) = \delta_{\alpha \beta} \log(r) - {r_\alpha
  r_\beta \over r^2} $ is often referred to as the 2D photon
propagator.  This formula represents the unsigned area of the loop
formed by $r(t)$.  It was used in \cite{cardy96:_geomet} to compute
areas formed by loops in a different context.

There is an intuitive way to understand why the unsigned area appears
in this context.  It can be computed geometrically as follows.  First
decompose a self-intersecting loop into a union of non
self-intersecting subloops (see Fig.~\ref{fig:combined}).  We can then
show that the variance of the dislocation charge enclosed by the loop
is equal to the sum of the unsigned areas of the subloops plus the sum
over all pairs of subloops of the areas of their intersections with a
plus sign if the two subloops are traversed in the same direction and
with the minus sign if they are traversed in opposite directions.

To simplify the task of averaging \rf{eq:unsign} over Brownian walks,
we follow the example of Ref.~\cite{cardy96:_geomet} and rewrite the
photon propagator in the following equivalent way
\begin{multline}
  \label{eq:unsignjohn}
  \overline{n^2}(R, T) = - {f \over 2} \int dt \int dt' \, \dot r_1(t)
  \, \dot r_1(t') \times \cr \delta(r_1(t) - r_1(t')) \ |r_2(t) -
  r_2(t')|.
\end{multline}
The advantage of this formula over \rf{eq:unsign} is in the fact that
$r_1$ and $r_2$ coordinates of the 2D Brownian walker decouple and
become two independent 1D Brownian walks.

It turns out that the average of \rf{eq:unsignjohn} over random walks
is logarithmically divergent at $t \approx t'$.  Anticipating that, we
only need to average \rf{eq:unsignjohn} at $t \rightarrow t'$. That
means, we can neglect all the terms in \rf{eq:vcor} which depend on
$T$ and $R_\alpha$, while keeping only the $\delta(t - t')$ term.  We
obtain
\begin{equation}
  \VEV{ |r_2(t) - r_2(t')| } \approx  \sqrt{2 D |t - t'| \over \pi},
\end{equation}
and
\begin{multline}
  \VEV{ \dot r_1(t) \dot r_1(t') \, \delta \left( r_1(t) - r_1(t')
    \right) } \approx \cr \sqrt{D \over 2 \pi \, |t - t'|} \left[
    \delta(t - t') - {1 \over 4 |t  - t'|} \right]. 
\end{multline}
Substituting this into \rf{eq:unsignjohn} we find the leading term
\begin{equation}
  \label{eq:div}
  \sigma(R, T) = \VEV{\overline{n^2}(R, T)} = {f D \over 8 \pi} \int
  dt \int dt'~{1 \over |t-t'|}.
\end{equation}
It is clear that the $t = t'$ divergence in \rf{eq:div} should be
regularized to give
\begin{equation}
  \label{eq:n^2_equal}
  \sigma(R, T) = {f D T \over 8 \pi} \log \left( {T \over \epsilon}
  \right).
\end{equation}
The regulator $\epsilon$ appears due to the fact that the trajectory
cannot wind around a given dislocation tighter than the dislocation
core size $a$.  The continuous formula \rf{eq:unsign} breaks down at
distances smaller than $a$.  The length $a$ corresponds to time
interval $\epsilon = a^2/D$ which is the average time it takes the
random walker to diffuse across a dislocation core.  We obtain
\begin{equation}
  \label{eq:answer}
  \sigma(R, T) = {f D T \over 8 \pi} \log \left( {D T\over a^2}
  \right).
\end{equation}
This is the second main result of this letter.  Notice that $\sigma(R,
T)$ does not depend on $R$, which should be contrasted with \rf{eq:nn}
which is valid when $f_+ \not = f_-$.

To summarize, we considered tracer diffusion in a layered system with
screw dislocations.  When the transverse diffusion coefficient of the
is small compared to the in-plane diffusion coefficient, tracer
particles are transported in the direction normal to the layers by
encircling screw dislocations.  We predict the shape of a cloud of the
tracer particles as a function of time.  We find that size of the
cloud in the direction normal to the layers (its height) grows faster
than its width.

To make quantitative predictions we need to address the following
concern.  The conventional transverse tracer diffusion coefficient
$D_\perp$ is never identically zero.  Tracer particles can be
transported along dislocation cores or point defects such as pores,
necks and passages as suggested by Constantin and Oswald in
\cite{constantin00:_diffus}.  They measured transverse diffusion in a
thin sample of lamellar phase of a surfactant/water mix.  Since their
sample contained only a few dislocations across its thickness, our
effect would therefore not be operative.  Instead one can estimate the
effect of an isolated screw dislocation using the classic result of
the statistics of winding numbers (e.g.\ \cite{drossel96:_winding}) to
be negligible compared with conventional diffusion $D_\perp$.

Going back to a layered system with many screw dislocations, we need
to estimate the time after which the superdiffusion due to
dislocations will dominate conventional transverse diffusion.  We
consider the case of equal densities of positive and negative
dislocations because unless there is a process at work which selects
dislocations of a certain charge, the difference in the densities will
be small.  The height of the cloud due to conventional diffusion is
roughly equal to the height due to superdiffusion when
\begin{equation}
  \label{eq:rough_estimate}
  D_\perp T \sim d^2 f D T \ln \frac{DT}{a^2},
\end{equation}
where $d$ is the interlayer distance.  Assuming that the dislocation
core size is equal to the interlayer spacing we obtain the crossover
time
\begin{equation}
  \label{eq:crossover}
  T_{\rm c} \sim \frac{d^2}{D} \, \exp\left({\frac{D_\perp/D}{d^2
        f}}\right).
\end{equation}
If this time is comparable to the experimentally available time, our
phenomenon should be observable.

Ref.~\cite{hamersky98:_anisotropy} measured the anisotropy of the
diffusion coefficient in clean samples of diblock copolymer to be
$D_\perp/D \approx 10^{-2}$.  Therefore, in order for our effect to
manifest itself, the defect density must be two orders of magnitude
larger than $d^2 f \approx 10^{-5}$ observed in the shear aligned
diblock copolymer system of Ref.~\cite{hamersky98:_anisotropy}.  The
diffusion of water mixed with egg phosphatidylcholine
\cite{wassall96:_pulsed_nmr} is even more anisotropic $D_\perp/D
\approx 10^{-3}$ so that our effect can be observed for smaller defect
densities.  A promising system is a mixture of lipid and surfactant
which undergoes a lamellar to nematic transition via proliferation of
screw dislocations \cite{dhez00:_lamel,constantin00:_diffus}.

To conclude we mention two phenomena which require a modification of
our predictions.  First, if screw dislocations are mobile, they will
tend to form bound dipole pairs of size comparable to the core size.
Since the bound pairs do not contribute to the transverse transport of
the tracer, only the density of free dislocations must be used in
Eq.~(\ref{eq:answer}).  In addition, the dislocation motion
\cite{holyst95:_disloc,blatter94:_vortic} will lead to an additional
mechanism for normal transport of the tracer.  Second, the presence of
edge dislocations impedes in-plane diffusion of the tracer.  This fact
may be successfully taken into account by renormalizing the in-plane
diffusivity.

The authors are grateful to R.~Selinger for seeding idea which led to
this calculation.

\bibliographystyle{prsty}

\begin{thebibliography}{10}

\bibitem{johnson00:_creep_tial_si}
D.~R. Johnson {\it et~al.}, Met. Mat. Trans. A {\bf 31},  2463  (2000).

\bibitem{wang98:_tial_ledges}
J.~N. Wang and T.~G. Nieh, Acta Mat. {\bf 46},  1887  (1998).

\bibitem{fredrickson96:_copolymers}
G.~H. Fredrickson and F.~S. Bates, Ann. Rev. Mat. Sci. {\bf 26},  501  (1996).

\bibitem{hamersky98:_anisotropy}
M.~W. Hamersky, M. Tirrell, and T.~P. Lodge, Langmuir {\bf 14},  6974  (1998).

\bibitem{wassall96:_pulsed_nmr}
S.~R. Wassall, Biophys. J. {\bf 71},  2724  (1996).

\bibitem{demenezes96:_study}
D.~E.~L. DeMenezes and E.~I.~V. Butler, Coll. Surf. B {\bf 6},  269  (1996).

\bibitem{postlethwaite94:_elect_cesium}
T.~A. Postlethwaite, E.~T. Samulski, and R.~W. Murray, Langmuir {\bf 10},  2064
   (1994).

\bibitem{allain86:_possib}
M. Allain, Europhys. Lett. {\bf 2},  597  (1986).

\bibitem{dhez00:_lamel}
O. Dhez {\it et~al.}, Euro. Phys. J. E {\bf 3},  377  (2000).

\bibitem{j.petermann79:_obser}
J.Petermann and R.~M. Gohil, Polymer {\bf 20},  596  (1979).

\bibitem{holyst95:_disloc}
R. Holyst and P. Oswald, Int. J. Mod. Phys. B {\bf 9},  1515  (1995).

\bibitem{bluestein01:_disloc_tgb}
I. Bluestein, R.~D. Kamien, and T.~C. Lubensky, Phys. Rev. E {\bf 63},  1702
  (2001).

\bibitem{spitzer58:_some}
F. Spitzer, Trans. Amer. Math. Soc. {\bf 87},  187  (1958).

\bibitem{pitman86:_asymp}
J. Pitman and M. Yor, Ann. Prob. {\bf 14},  733  (1986).

\bibitem{drossel96:_winding}
B. Drossel and M. Kardar, Phys. Rev. E {\bf 53},  5861  (1996).

\bibitem{kholodenko98:_elemen_brown}
A.~L. Kholodenko, Phys. Rev. E {\bf 58},  R5213  (1998).

\bibitem{cardy96:_geomet}
J. Cardy,  in {\em Fluctuating geometries in statistical mechanics and field
  theory}, {\em Les Houches Summer School Proceedings}, edited by F. David, P.
  Ginsparg, and J. Zinn-Justin (Elsevier, Amsterdam, 1996), .

\bibitem{constantin00:_diffus}
D. Constantin and P. Oswald, Phys. Rev. Lett. {\bf 85},  4297  (2000).

\bibitem{blatter94:_vortic}
G. Blatter {\it et~al.}, Rev. Mod. Phys. {\bf 66},  1125  (1994).

\end{thebibliography}

\end{document}